\documentclass[11pt, a4paper]{article}
\usepackage{fullpage}
\usepackage{amsfonts}
\usepackage{amssymb}
\usepackage{amsmath}
\usepackage{amsthm}
\usepackage{graphicx}
\usepackage{bm}
\usepackage[makeroom]{cancel}
\usepackage{enumitem}
\usepackage{url}
\usepackage[margin=.9in]{geometry}
\usepackage{amsopn}
\usepackage{mathtools}
\usepackage{hyperref}
\usepackage{doi}
\usepackage{cite}
\usepackage{overpic}
\usepackage[symbol]{footmisc}
\usepackage{booktabs}
\usepackage{siunitx}
\usepackage{wasysym}

\usepackage{listings}
\usepackage{xcolor}
\definecolor{codegreen}{rgb}{0,0.6,0}
\definecolor{codegray}{rgb}{0.5,0.5,0.5}
\definecolor{codepurple}{rgb}{0.58,0,0.82}
\definecolor{backcolour}{rgb}{0.95,0.95,0.92}
\lstdefinestyle{mystyle}{
    backgroundcolor=\color{backcolour},   
    commentstyle=\color{codegreen},
    keywordstyle=\color{magenta},
    numberstyle=\tiny\color{codegray},
    stringstyle=\color{codepurple},
    basicstyle=\ttfamily\normalsize,
    breakatwhitespace=false,         
    breaklines=true,                 
    captionpos=b,                    
    keepspaces=true,                 
    numbers=left,                    
    numbersep=3pt,                  
    showspaces=false,                
    showstringspaces=false,
    showtabs=false,                  
    tabsize=4,
}
\begin{document}
\begin{center}
    \Large \bf An Interpretable Data-Driven Model of the\\ Flight Dynamics of Hawks
    % Interpretable dynamic structure of hawk flight
\end{center}
\begin{center}
    Lydia France$^{1,2}$\footnote[1]{Corresponding authors (lydia.france@biology.ox.ac.uk)},
    Karl Lapo$^{3}$,
     J. Nathan Kutz$^{4}$
\end{center}
\begin{center}
    \scriptsize{
    ${}^1$ Department of Biology, University of Oxford, Oxford, UK  \\ 
    ${}^2$ The Alan Turing Institute, London, UK  \\ 
    ${}^3$ Department of Cyrospheric and Atmospheric Sciences, University of Innsbruck, Innsbruck, Austria \\ 
%    ${}^4$ Department of Electrical and Computer Engineering, University of Washington, Seattle, WA 98195, United States \\
%     ${}^5$ Department of Applied Mathematics, University of Washington, Seattle, WA 98195, United States     }
     ${}^4$ Autodesk Research, 6 Agar Street, London, UK     }
\end{center}

\begin{abstract}
Despite significant analysis of bird flight, generative physics models for flight dynamics do not currently exist. Yet the underlying mechanisms responsible for various flight manoeuvres are important for understanding how agile flight can be accomplished. Even in a simple flight, multiple objectives are at play, complicating analysis of the overall flight mechanism. Using the data-driven method of {\em dynamic mode decomposition} (DMD) on motion capture recordings of hawks, we show that multiple behavioral states such as flapping, turning, landing, and gliding, can be modeled by simple and interpretable modal structures (i.e. the underlying wing-tail shape) which can be {\em linearly} combined to reproduce the experimental flight observations.  Moreover, the DMD model can be used to extrapolate naturalistic flapping. Flight is highly individual, with differences in style across the hawks, but we find they share a common set of dynamic modes.  The DMD model is a direct fit to data, unlike traditional models constructed from physics principles which can rarely be tested on real data and whose assumptions are typically invalid in real flight.  The DMD approach gives a highly accurate reconstruction of the flight dynamics with only three parameters needed to characterize flapping, and a fourth to integrate turning manoeuvres.  The DMD analysis further shows that the underlying mechanism of flight, much like simplest walking models, displays a parametric coupling between dominant modes suggesting efficiency for locomotion.  
%
%
%Flying is so easy, even a bird can do it.
%- interpretable
%- simple
%- data-driven
%- mixture of modal dynamics for multiple flight behaviors
%- patterns hold across birds and across flights
%- variations and flight styles
%- doubling mode which is like walking
% - Across eight individuals, hawks use a shared, low-dimensional morphing dynamical structure.
\end{abstract}

\section{Introduction}

% P1:  overview of bird flight... hard plus multi-objective and variation. control mechanism. So much context. 

% P2:  Data-driven models are emerging... build models directly from data versus derivations. Flow physics, assumptions. Learning physics from data. Some of these are  interpretable, DMD is one of these and we use it.

% P3: Bio-informed, naturalistic flight, how to extract data from videos so you don't have to use dead birds. Highlight your PCA paper.

% P4:  We show!  DMD on data and highlight results.  Dynamics.

% P5: Overview of sections ????

% One of our main points: flying MUST be easy
% It is a hypothesis
% We also need a way to explain what the mode doing

%\subsection{Context of Bird Flight}

Birds execute flight maneuvers with complex and intricate wing-tail shape changes \cite{chin2016a, harvey2022d}. Across a single trajectory, a bird may combine flapping, gliding, braking, obstacle avoidance and precise landing all within the time-frame of seconds \cite{France2025, perching_2022}. These behaviors arise from high-dimensional morphing, the continuous wing-tail shape adjustments that produce inertial and aerodynamic effects \cite{taylor2002, altshuler2015a}. Bird flight is also highly individual, with variable flight techniques even among related birds or across life history \cite{France2025, perching_2022, martinez_groves-raines_steady_2024, dakin_individual_2020}. Despite extensive biomechanical study, how wing-tail morphing are organized during natural, multi-objective flight is unknown, limiting our understanding of flight control. 

Classic flapping-flight models typically adopt a bottom-up approach, where kinematics are prescribed, for example a simplified wing-beat or joint-angle trajectory \cite{pennycuick1968a, pennycuick2008}. Kinematics are used as inputs and aerodynamic forces are then computed with quasi-steady, unsteady, or computational fluid dynamics (CFD) models \cite{ellington1984, tobalske2022b, ellington1996, shen2025, song2022a}. Pennycuick's power decomposition, Rayner's vortex-ring theory, and modern CFD have successfully explained how specific anatomical features (wing-aspect ratio, wing twist, body mass) relate to aerodynamic performance \cite{pennycuick1968a, rayner1979, rayner2001, usherwood2002, tobalske2022b}. However, such models are limited in explaining natural animal flight. Firstly, temporal dynamics remain absent from most models. The rate and trajectory of morphing, not solely the final configuration, critically determine flight control \cite{tobalske2022b, wang2000}. Additionally, these models are rarely validated against real flight data, and their underlying assumptions, such as the quasi-state hypothesis, are typically invalid under naturalistic conditions \cite{ellington1984, chin2019b}. Natural flight involves multiple objectives, such as maintaining speed and lift while avoiding an obstacle, as well as transitioning between a continuum of flight behaviors such as gliding and flapping and landing. To understand the complex flight dynamics of natural bird flight, a bottom up approach has thus far proven to be unsuitable. 

% Instead we invert the classical approach. Rather than presuming aerodynamic mechanisms, we use a data driven approach to discover the underlying dynamical structure directly from naturalistic flight recordings. Dynamic Mode Decomposition identifies the simplest model basis that can reconstruct the measured hawk wing-tail morphing \cite{dmd_book}. From this empirical approach, we provide an interpretable and compact model of morphing bird flight, which can be used to generate realistic behavior sequences. 

% Even high fidelity CFD models remain fundamentally conditioned on recorded or idealized kinematics as input, rather than discovering the underlying dynamical structure that generates morphing itself. Recent work demonstrates that the rate and trajectory of morphing, not solely the final configuration, critically determine flight control, yet these temporal dynamics remain absent from most models. As a result, we lack a compact, interpretable model of multi-objective morphing flight, and the ability to generative prediction of realistic behavioral sequences.

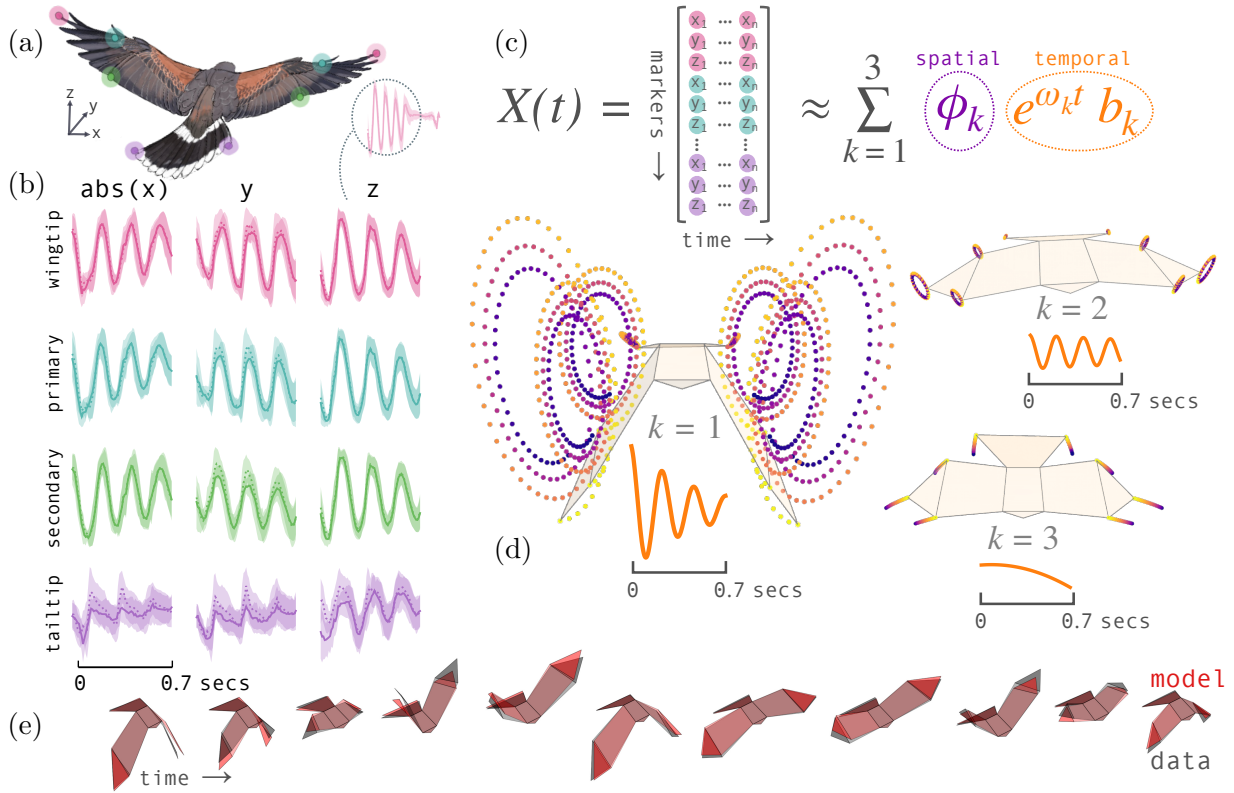
\begin{figure}[t]
\centering
\begin{overpic}[width=0.95\textwidth]{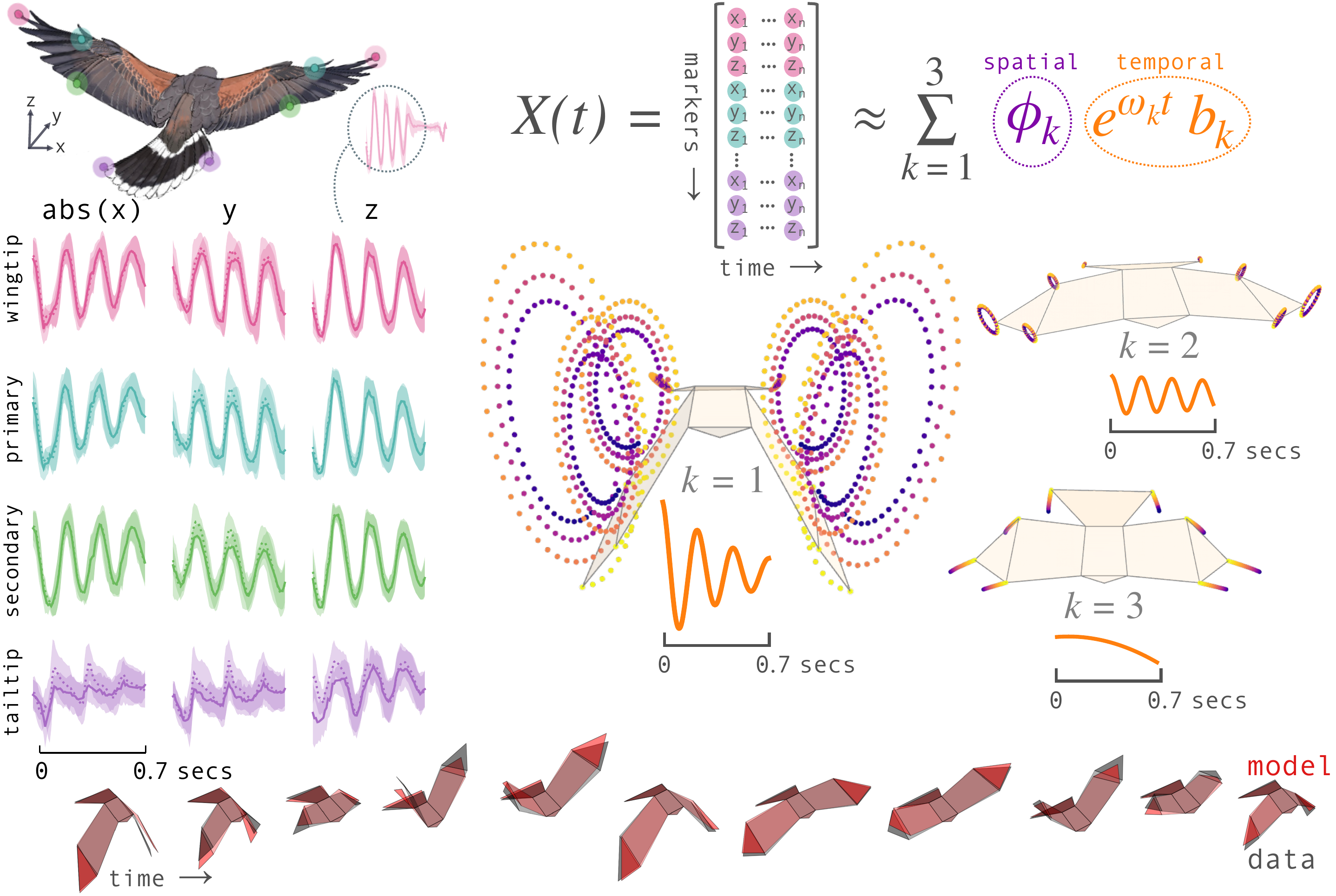}
\put(-3,63){(a)}
\put(-3,51){(b)}
\put(38,63){(c)}
\put(38,20){(d)}
\put(-3,5){(e)}
\end{overpic}  
\caption{Dynamic Mode Decomposition on hawk flapping flight reproduced with a model with 3 terms. (a) A reconstruction of eight wing and tail landmarks in 3D of a Harris' hawk flying between two perches using motion capture.  (b) Restricting data to just flapping behaviour lasting 0.7s after take-off, a binned mean was taken from 67 flights by the same individual (bin size = 0.005s, shaded area ±1 standard deviation). (c) Using DMD, we approximate the flapping behaviour as the sum of three dynamic modes, which contain both a spatial, $\phi_k$, and temporal $e^{\omega_k t}$ component. (d) Projecting the effect of each mode onto the hawk shape, with time going from yellow to purple: the first mode (k(1) freq = 4.51 Hz) contains the circular wingbeat oscillation with a gradual decay; the second mode shows a close to double frequency of the first (k(2) freq = 8.76 Hz) showing a small oscillation on top of the main wingbeat; the third mode shows a gradual increase in wingspan and tail dropping (k(3) freq= 0.00Hz).  (e) From the flapping flight, three dynamic modes combine to reconstruct the original flapping behaviour with total RMSE error within 1.2\% the hawk's maximum wingspan (total absolute error = 12mm).}
\label{fig:DMDSummary}
\end{figure}

Recent advances in data driven models provide an alternative modeling strategy, where structure and dynamics are learned from the behavioral data directly~\cite{data_book}. {\em Dynamic mode decomposition} (DMD)~\cite{schmid_2008,schmid_2010} is defined as an algorithm that generates a linear decomposition.  But unlike standard PCA or SVD methods~\cite{kutz2013data}, it has the added benefit of correlating temporal and spatial modes~\cite{dmd_book}. We use wing and tail measurements of hawks flying between two perches, with and without an obstacle, and find the shared dynamic modes across individuals and during different manoeuvres. Unlike black box deep learning models that are susceptible to overfitting, we build a low-dimensional subspace with interpretable linear dynamical structure. As illustrated in Fig.~\ref{fig:DMDSummary}, we show the low-dimensional dynamic structure of hawk morphing flight using DMD. Previous work identified the low-dimensional spatial structure of wing–tail geometry during morphing, here we focus on the dynamics and how hawks move through this morphing space during natural flight using a small number of interpretable modal features, or poses. We find despite context- and individual-dependent flight techniques, the hawks share common dynamic modes. DMD deconstructs the multi-objective flight behaviors into linearly superimposed dynamics, providing an interpretable and simple model for hawk morphing flight. While showing high reconstruction accuracy of natural flight behaviors, the model can also extrapolate forwards and generate natural flight behaviors. Our data-driven model therefore emerges directly from real bird flight data, but with interpretable and generalizable results. 

Importantly, the DMD analysis of the data reveals the key features of bird flight.  These include dominant poses, or wing-tail configurations, that evolve (oscillate) in time in order to reproduce the observed flight data, as shown in Fig.~\ref{fig:DMDSummary}.  And despite the seeming complexity of flight, the DMD algorithm reveals that flight dynamics can be parametrized by three parameters relating to the linear addition of three key wing-tail body shapes.  These critical shapes driving flight dynamics are the DMD modes.
The DMD analysis serves at least two purposes.  For the biologist, it represents an analysis architecture for real data recordings that is capable of extracting simple, biologically interpretable outcomes for understanding the dominant components of biolocomotion. For the engineering, the DMD regression provides a theoretical framework for the construction of biolocomotion models, thus allowing for engineering proxies of biology (robotics and simulation models) and critical insight into potential control algorithms potentially executed by birds for control of flight.  Such insights are completely data-driven and do not require recourse to the complex kinematic or aerodynamic models that exist in the literature.

\section{Methods}

\subsection{Data and Flights}

%\subsection{Methods: Bird Flight Experiments}
Experiments used five Harris's hawks (\textit{Parabuteo unicinctus}, four males, one female), with varying experience and age. The hawks were flown between two perches for a food reward under motion capture cameras and wearing retro-reflective markers on their feathers (see Figure \ref{fig:DMDSummary}). The full experimental methodology and statistical analysis of morphing flights from all 1659 flights are available in previous work \cite{France2025}. Previous work has focused on the whole body trajectories and the shape changes involved in morphing flight, here we for the first time formally analyze the motion of the hawks' wings and tail over time. We separated the flight recordings by perch-perch distance, by individual, and, where present, left or right turns around an obstacle. As the flights contain different behaviors, for analysis we also separated the flights into flapping (0.7s from the take-off jump to a final wingbeat), gliding (from the end of the final wingbeat to the touch of the destination perch), or treated as a full flight. Within each individual and for each experimental period, we used a binned mean over time (bin size 0.005s) for each of the eight 3D feather marker positions relative to the center of the bird (shown in the upper left of Fig. \ref{fig:DMDSummary}). The whole body rotations of the hawks in flight are included in the measurements. In Figures \ref{fig:DMDSummary} and \ref{fig:turning}, we show $\pm1$ standard deviation as shading around the binned mean. For DMD, we fit to the average flight per individual and experiment. DMD was also fit to every flight sequence subject to a quality filter: minimum 30 frames, and no gaps for more than 0.15s, and the first measurement in the flight within 0.2s of takeoff.

\subsection{Dynamic Mode Decomposition (DMD)}
%\subsection{Dynamic Mode Decomposition}

%What do we mean by a mode? -> We mean a pose and how it changes with time.

%Use DMD as a way of identifying coherent dynamics within complex biomechanical systems.

DMD seeks to decompose time-varying data sets into a low-rank coherent set of spatiotemporal structures~\cite{schmid_2010, tu_2014, dmd_book, schmid2022dynamic}.  Specifically, the temporal evolution is assumed to be approximated by a linear dynamical system $\dot{\mathbf{x}}(t) = \mathbf{A} \mathbf{x}(t)$ whose evolution dynamics are exponentials of the form
\begin{equation}
    \mathbf{\tilde{x}}(t) = \sum_{j=1}^N b_j \boldsymbol{\phi}_j \exp(\omega_j t)
    \label{eq:DMDsolution}
\end{equation}
with the imaginary component of the exponent $\omega_j$ modeling periodic temporal oscillations and the real component modeling exponentially growing or decaying phenomena.   The $\mathbf{\tilde{x}}$ refers to an approximation of the original data, $\mathbf{x}$. The $\boldsymbol{\phi}_j$ are the DMD spatial modes. For the data considered here, these will be the wing-tail body shapes that determine the leading behavior in flight. The parameter $b_j$ is the weight of each DMD body shape when linearly superimposing for reconstructing the flight data. Finally, $j$ specifies the rank of a mode, up to a total rank of $N$. A DMD mode consists of a conjugate pair of both $\phi_j$ and $\omega_j$. Thus, in our application of DMD we discover wing-tail body shapes that move together with a specific frequency.

Regressing to the solution form (\ref{eq:DMDsolution}) through an optimized DMD~\cite{optdmd,bopdmd, pydmd} algorithm has many benefits, and it allows for tasks such as model reduction and future-state prediction. This insight paired with the equation-free nature of the algorithm has led to the widespread application of DMD across many scientific fields, including fluid dynamics \cite{schmid_2008, schmid_2009, schmid_2010, Noack2016jfm}, epidemiology \cite{proctor_2015}, neuroscience \cite{brunton_2016, alfatlawi_2020}, finance \cite{mann_2016}, robotics \cite{berger_2015, abraham_2019, bruder_2019}, plasma physics \cite{taylor_2018, kaptanoglu_2020} and control~\cite{dmd_book} to name a few. The method has additionally been connected to Koopman operator theory and has since become the standard approach for approximating the Koopman operator from data \cite{rowley_2009, tu_2014, modern_koopman}.

Over the years since the algorithm's conception, the DMD algorithm has seen a great deal of improvements, variants, and extensions \cite{schmid2022dynamic}, with one of the most notable innovations being the optimized DMD algorithm \cite{optdmd, pydmd}. Optimized DMD and its more recent successors \cite{bopdmd, robust_dmd, optdmd_c} generally utilize variable projection in conjunction with other techniques from optimization in order to compute the spatiotemporal DMD modes. This allows optimized DMD to handle unevenly-sampled snapshots and produce results that are significantly more robust to measurement noise compared to previous approaches like exact DMD \cite{tu_2014}. The variable projection technique extends to generalized objective function formulations for obtaining the components of \eqref{eq:DMDsolution}, enabling  robust DMD \cite{robust_dmd}, which may include model regularizers, constraints, and robust data-fitting penalty functions like the Huber penalty. This flexibility, paired with fast and reliable optimization algorithms have made optimized DMD and its robust extensions highly applicable to real-world applications.

\subsection{Rank Choice}

The choice of DMD rank (the parameter $N$ in (\ref{eq:DMDsolution})) represents a trade-off between accurately capturing the observed dynamics and avoiding over-parameterization. While higher-rank decompositions can accommodate increasingly subtle features of the data, they may also introduce modes that reflect redundancy rather than independent physical structure. For this analysis, we initially fit either four or six modes (two or three complex-conjugate pairs) to the data. This choice was guided by a combination of quantitative diagnostics, including variance explained, total reconstruction error (RMSE), amplitude ranking, and inspection of the reconstructed modes and their temporal signatures. Across all conditions, these diagnostics indicated progressively smaller improvements in reconstruction quality as additional modes were added.

The final truncation was informed by whether individual modes fell into an exclusion category indicative of overfitting. Specifically, modes were excluded if they (i) primarily counteracted or canceled the deformation introduced by a dominant oscillatory mode, (ii) influenced only a single wingbeat or a highly localized temporal segment, or (iii) appeared to split a single coherent oscillatory pattern into multiple closely related modes with slightly different frequencies. Such behavior reflects redundancy inherent to non-orthogonal modal decompositions rather than distinct, repeatable kinematic dynamics. For reasons of parsimony and physical interpretability, only modes exhibiting coherent structure across the flight sequence were retained. We therefore retained 3 mode pairs for the flapping only data and 1 pair for gliding. For the full sequence we retained 3 pairs for flights with an obstacle and 2 for flights without. 

\subsection{Marker and Mode Reconstruction}
The accuracy of the DMD reconstructions was assessed using the combined reconstruction from \eqref{eq:DMDsolution} for the selected $N$ modes compared with the original marker positions throughout a flight. Error was quantified using root mean square error (RMSE) and  computed over time, by marker, and both. Including both the oscillatory (imaginary part of $\omega_k$) and the exponential growth or decay (real part of $\omega_k$) provides the closest fit to the observed data. 

To visualize the contribution of individual modes, we reconstructed the original signal with a single DMD mode or single complex conjugate pair from \eqref{eq:DMDsolution}. Reconstructing individual modes facilitates interpretation of how the mode contributes to the overall kinematics. A single physical mode consists of a conjugate pair of DMD modes, denoted $P$, which correspond to a single, coherent spatial structure $|\phi_k|$ with temporal deformation given by an oscillatory frequency given dictated by $Im(\omega_k)$ and growth or decay rate dictated by $Re(\omega_k)$ \cite{lapoPhasorNotationDynamic2025}. Modes without an imaginary component ($Im(\omega_k) = 0$) are treated as self-conjugate and contribute only once without a corresponding conjugate pair. 

The modes can be used for forecasting reconstructions forwards in time, therefore providing a generative model of hawk flight. In such models, exponential growth or decay is undesirable, as it leads to non-physical divergence or collapse of motion. We therefore reconstruct selected conjugate mode pairs while explicitly removing the growth/decay term:

\begin{equation}
    \mathbf{\tilde{x}}_{k,stable}(t) \approx \sum\limits_{k\in P}  2b_k |\phi_k|\cos (\omega_kt - \theta_k)
    \label{eq:DMD_stable}
\end{equation}

where $P$ denotes a subset of conjugate pairs. The formulation in \eqref{eq:DMD_stable} retains the spatial structure, frequency, and phase relationships of each oscillatory mode while enforcing constant amplitude over time. For reconstructing flight beyond the original time window, the kinematics are stable and periodic. 

All reconstructions describe deviations about a mean pose. For absolute marker trajectories, the time averaged shape $\bar{x}$ was added back to the reconstructed displacements:

$$x_\text{abs}(t) = \bar{x} + x(t)$$
This provides interpretable marker trajectories for animation or traces suitable for comparison with the original measured data.

\section{Results and Discussion}

%%---- Explaining the modes -----
\subsection{DMD Modes for Flapping Flight}

DMD represents the observed kinematics as the sum of a small number of modes, each of which is a characteristic pattern of coordinated motion of the wings and tail over time. Taking an averaged flapping flight sequence (Figure~\ref{fig:DMDSummary}) for 0.7 seconds after take-off from a perch, three modes are sufficient to reconstruct the original kinematics with high accuracy. Each mode has a spatial pattern (the shape changes of wings and tail) and a temporal evolution (its oscillation frequency and how its amplitude changes over the trajectory) which are depicted in Figures \ref{fig:DMDSummary}d and \ref{fig:individuals}.\\

\noindent
{\bf DMD Mode One:}
The first mode captures the wingbeat. It describes the large-amplitude, generally circular stroke of the wings, the associated vertical motion of the tail, and the folding of the hand-wing during the tip-reversal at the end of each downstroke. This mode oscillates at the wingbeat frequency (4.5 Hz), and its amplitude decays over successive beats, indicating a gradual reduction in wingbeat amplitude as the hawk transitions away from powered take-off.\\

\noindent
{\bf DMD Mode Two:}
The second mode is a smaller oscillation superimposed on the wingbeat mode. This mode involves modest circular motion of the wings in the opposite direction to the wingbeat mode and without tail motion, and it oscillates at approximately twice the flap frequency. This doubled-frequency component modulates the shape of each wingbeat, adding finer-scale adjustments within a single down–up cycle.  The frequency of mode two is approximately double that of mode one.\\

\noindent
{\bf DMD Mode Three:}
The third mode is predominantly a non-oscillatory change in posture. It corresponds to an increasingly outstretched configuration: the wingspan and tail span increase steadily over time, and the tail drops. This mode captures the gradual shift from a power-generating behavior toward the extended posture characteristic of gliding.\\

When summed, these modes show how the hawks transition from take-off acceleration using flapping flight to the beginning of gliding. While wingbeat amplitude decreases, the hawk simultaneously increases wingspan and tail span, revealing a coordinated transition from high-amplitude flapping at take-off to a more extended, glide-oriented posture. Previous work has found flight behaviors exist on a continuum, which is supported here \cite{France2025}. Gliding does not "switch on", instead is the increasing contribution of a dynamic mode in favor of the wing-beat mode.

\subsection{Consistency across Individuals}

\begin{figure}[t]
  \centering
  \includegraphics[width=0.95\textwidth]{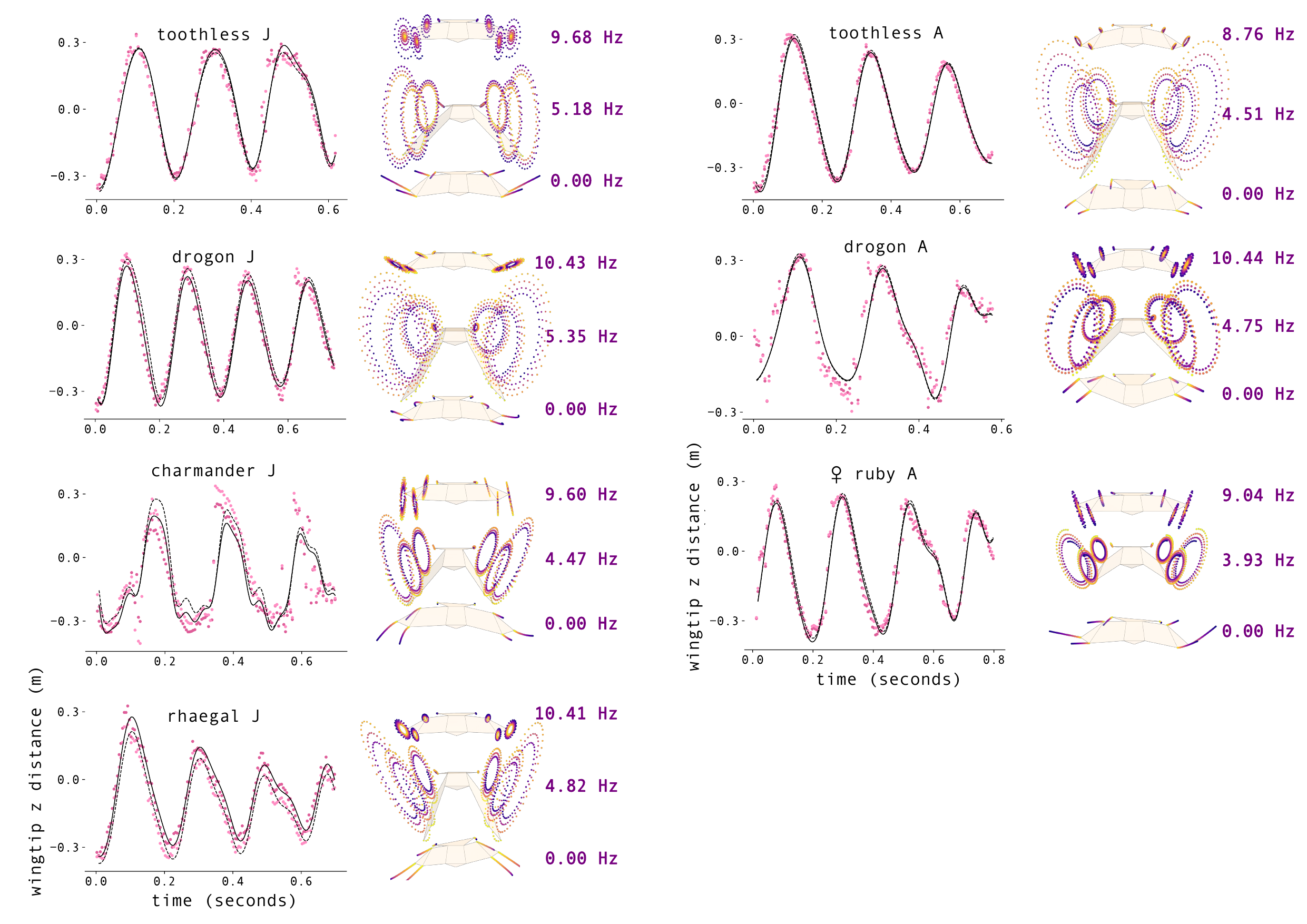}
  \caption{DMD fit to 9m straight flights by juvenile (J) and adult (A) hawks' flapping behaviour for 0.7s after take-off from a perch. Left plots: binned mean wingtip positions relative to the center of mass in z (pink: left wing, red: right wing) overlaid with the DMD model fit (black: right wing, black dashed: left wing). Right plots: the projection traces show the influence of each dynamic mode on the wings and tail shape. Individual variation in flight is high, with bilateral asymmetry, and the same hawk showing a change in technique after three years of maturation. Despite individual variations, the three modes show similar spatial patterns. The dominant mode showed the circular wingbeat with folding, and another mode close to double this frequency showing modulation of the wings on top of the wingbeat. A slower, non-oscillating mode also showed an increase in wing and tail area, though achieved in different ways.}
  \label{fig:individuals}
\end{figure}

We repeated the DMD analysis on binned averaged flapping flight data from seven individual hawks, including two individuals measured both as juveniles and again as adults approximately three years later. Flights were a total of 9m between two perches, with 0.005s bins and truncated to the end of flapping. Across all individuals, the qualitative structure of the decomposition was consistent (Figure \ref{fig:individuals}). In every case, the leading components consisted of (i) a slowly varying mode corresponding to the average pose with increasing wingspan/tail-span, (ii) a dominant flapping mode at the wingbeat frequency, and (iii) a secondary oscillatory mode with frequency close to twice the wingbeat frequency. 

Although the qualitative mode structure was shared across the hawks, the precise frequencies associated with the oscillatory modes varied across individuals and developmental stage (juvenile vs adult). Wingbeat frequencies ranged from approximately 3.93–5.35 Hz across individuals, with corresponding secondary modes occurring near twice this frequency. Juvenile flights generally exhibited higher wingbeat frequencies than adult flights from the same individual, as expected from scaling of flight kinematics with body size and development. Despite these variations, the ratio between the primary and secondary oscillatory frequencies were close to 2:1. The non-oscillating mode also showed individual variation, though with a shared outcome of increasing wing and tail area over the course of the flapping flight.

\begin{table}[t]
\centering
\caption{Dominant DMD frequencies for individual birds and the ratio between the two dominant modes. Across individuals, the frequency ratio is approximately two.}
\label{ta:freq}
\begin{tabular}{ll
                S[table-format=2.2]
                S[table-format=1.2]
                S[table-format=1.2]}
\toprule
\textbf{Individual} & \textbf{Stage} &
{$f_1$ (Hz)} & {$f_2$ (Hz)} & {$f_2/f_1$} \\
\midrule
Toothless  & Juvenile &  9.68 & 5.18 & 1.87 \\
           & Adult    &  8.76 & 4.51 & 1.94 \\
Drogon     & Juvenile & 10.43 & 5.35 & 1.95 \\
           & Adult    & 10.44 & 4.75 & 2.20 \\
Charmander & Juvenile &  9.60 & 4.47 & 2.15 \\
Ruby (\venus)   & Adult    &  9.04 & 3.93 & 2.30 \\
Rhaegal    & Juvenile & 10.41 & 4.82 & 2.15 \\
\bottomrule
\end{tabular}
\end{table}

\subsection{Double Frequency Mode}

We further investigated the low-amplitude mode that oscillates at close to twice the wingbeat frequency as characterized in Table~\ref{ta:freq}. Reconstruction of the mode shows the deformation is localized to the distal wing, amplitude increasing towards the wingtip and negligible involvement of the tail. The presence of this mode was consistent to all the individuals during flapping flight though with individual variation (Figure \ref{fig:individuals}). The kinematic signature of this mode was non-rigid. The distances between the markers do not remain constant and instead varied $\approx0.12-0.24$m. The spatial pattern was also strongly symmetric across the left and right wings. Temporally, the mode is closely phase-locked to the wingbeat cycle with its peak contribution occurred at the same point in the wingbeat across the flapping flight. Removal of this mode from the reconstruction reduced the integrated wingtip velocity by approximately 5\%, demonstrating a measurable aerodynamic effect despite its small energy content. 

One possible explanation for this mode is passive aeroelastic deformation of the feathers. However, this interpretation is weakened by the spatial structure of the mode, where large contributions arise from a marker in the middle of the primary feathers (blue marker in Figure \ref{fig:DMDSummary}) and almost none from the marker on the tip of the long tail feather (pink) which would also show passive bending. An alternative and more plausible explanation is that this mode reflects active musculoskeletal control. Joint-angle measurements in pigeons show a clear double-peaked pattern within each wingbeat cycle \cite{robertson2012}. The doubled-frequency DMD mode shares several key features with distal forearm muscle activity reported in pigeon electromyography studies: a near-harmonic frequency relative to the wingbeat, consistent within-cycle timing when referenced to the half-stroke, left–right symmetry, and spatial confinement to the distal wing. In particular, the pronator superficialis and supinator muscles in pigeons exhibit alternating bursts within each wingbeat that rotate the forearm during downstroke and upstroke sub-phases. Such periodic forearm rotation produces distal torsion of the wing, acting as a modulatory mechanism that fine-tunes aerodynamic forces within each stroke rather than contributing directly to stroke generation.

This interpretation is consistent with the kinematic signature of the mode: the elliptical wingtip motion is oriented in the opposite direction to the primary wingbeat mode, indicative of torsional or rotational deformation rather than translational flapping. Taken together, these observations support the interpretation that this DMD mode captures active distal wing modulation occurring twice per wingbeat, likely the forearm twisting driven by musculoskeletal activation rather than passive aeroelastic effects.

\subsection{Full Flight DMD Modes}

Fitting DMD to the full flight trajectory (take-off, flapping, gliding, landing) shows the complexity of different flight behaviors as the varying contributions and amplitudes of dynamic modes over the flight trajectory. While DMD has weaker fit when applied to very different dynamics \cite{lapoMethodUnsupervisedLearning2025}, it is useful for finding dynamics that contribute throughout the flight trajectory. As before, the first mode describes the wingbeat frequency but attenuating further. The wingspan increasing mode then also includes the dramatic whole body rotation known as the rapid pitch-up maneuver.  With DMD, this well-known highly coordinated pattern of morphing and rotation emerges from the data. Rather than ``switching on", the transition to gliding is the increasing contribution of one mode over another; the wing-beat mode attenuates and then the increasing span mode dominates.

\subsection{Turning Flights}
\begin{figure}[t]
  \centering
  \includegraphics[width=\textwidth]{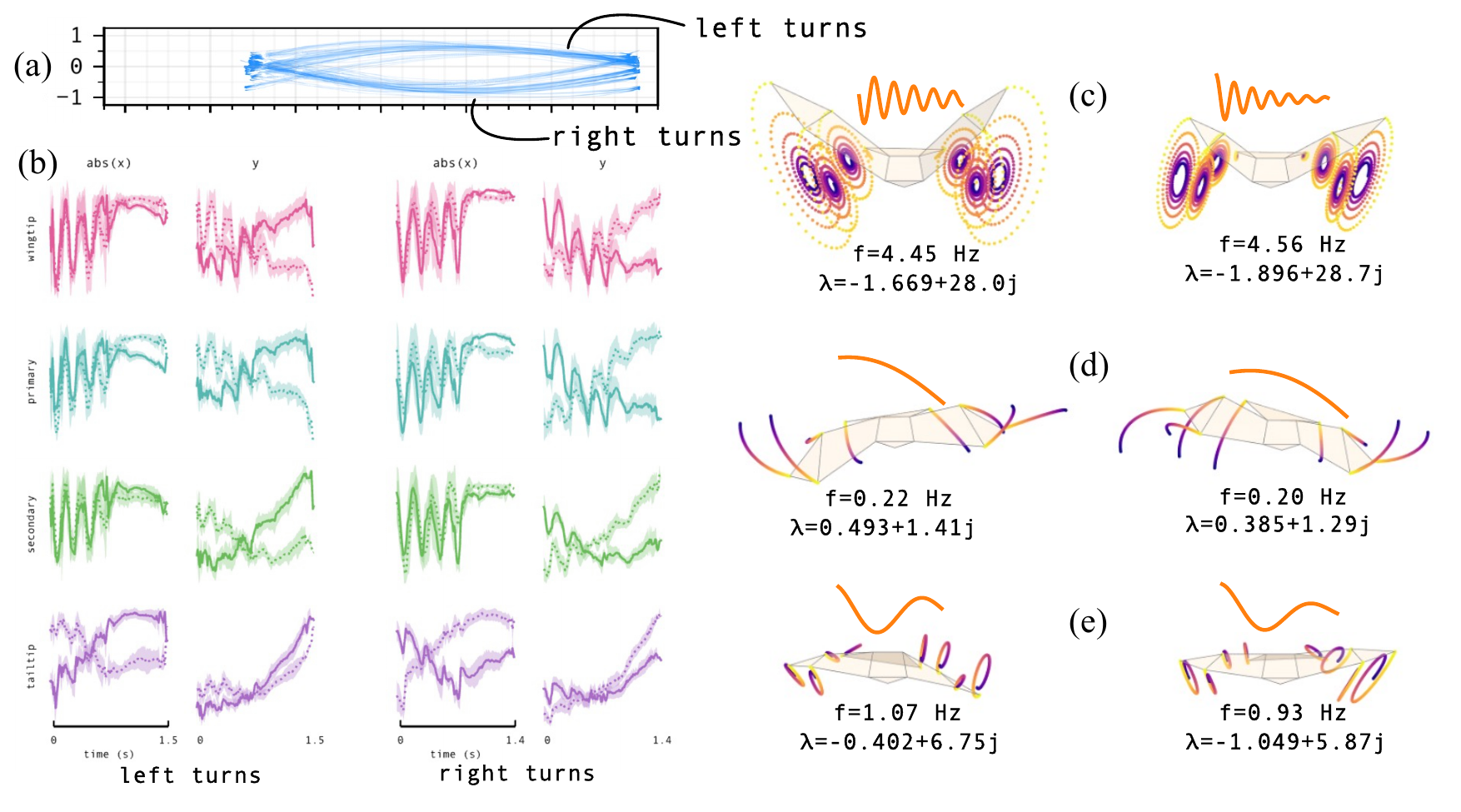}
  \caption{Turning modes. (a) 9m flights between two perches with a midpoint obstacle were recorded by one hawk (Toothless), (b) with marker data split into left (36 flights) and right (30 flights) and a binned mean taken at every 5mm. As before, the modes describe the (c) flapping wingbeat oscillation, and (d) an increase in wingspan and pitching up, but the DMD modes now include significant banking rotation, which are mirrored for left and right turns. (e) An asymmetric mode creates left-right imbalances likely driving the turn.}
  \label{fig:turning}
\end{figure}

We fit DMD to the marker positions on hawks flying 9m flights between two perches with an obstacle at the midpoint (Fig.~\ref{fig:turning}a). We split the flights into left or right turns, as the hawks were free to chose their path to the perch. Using a binned mean across the flights and split by individual  (Fig.~\ref{fig:turning}b), we fit six DMD modes. As before, we found a flapping frequency mode, though this time with the addition of a banked whole body rotation (Fig.~\ref{fig:turning}c). The next mode pair also described the wing and tail span increasing over time with the rapid pitch-up for landing, and again was also banked (Fig.~\ref{fig:turning}d). We show the banking rotation of these modes mirror each other depending on left or right turn. An additional mode pair was found on top of the flapping and gliding modes (Fig.~\ref{fig:turning}e). With a lower frequency of around 1Hz, this mode shows a spatially asymmetrical effect for the left versus right wings and sides of the tail. It therefore likely describes the asymmetrical input from the hawk to enable the turn and recovery around the obstacle before landing. This mode showed high individual variation. We also found variation within the same individual between left and right turns, which would be expected in natural flights where an animal may show dominance or preference for one side (i.e. ``left or right winged"). These results further support findings from previous papers that there is no one stereotyped method for maneuvering flight and variation can exist between contexts, individuals, or instances  \cite{martinez_groves-raines_steady_2024, dakin_individual_2020, France2025}. These are also a key example of multi-objective flight, where forward acceleration, transition from flapping to gliding, the landing maneuver, as well as inducing a turn all occur within 1.5 seconds from take-off. The combination of these dynamics are described accurately by the sum of three dynamic mode pairs.

The DMD fit to an entire flight has lower fidelity decomposition of the flapping phase due to fitting multiple discrete phases of dynamics \cite{lapoMethodUnsupervisedLearning2025}. It is for this reason we do not recover the doubled frequency mode found in the previous experiments. The doubled frequency mode on top of the main wingbeat frequency emerges when fitting DMD with higher rank, but with a penalty of transient or over-fitted modes. Thus, the doubled frequency mode is excluded in order to not include potentially non-physical modes and so on balance three pairs are suitable for finding modes that occur throughout the flight.

\subsection{Reconstruction Accuracy}

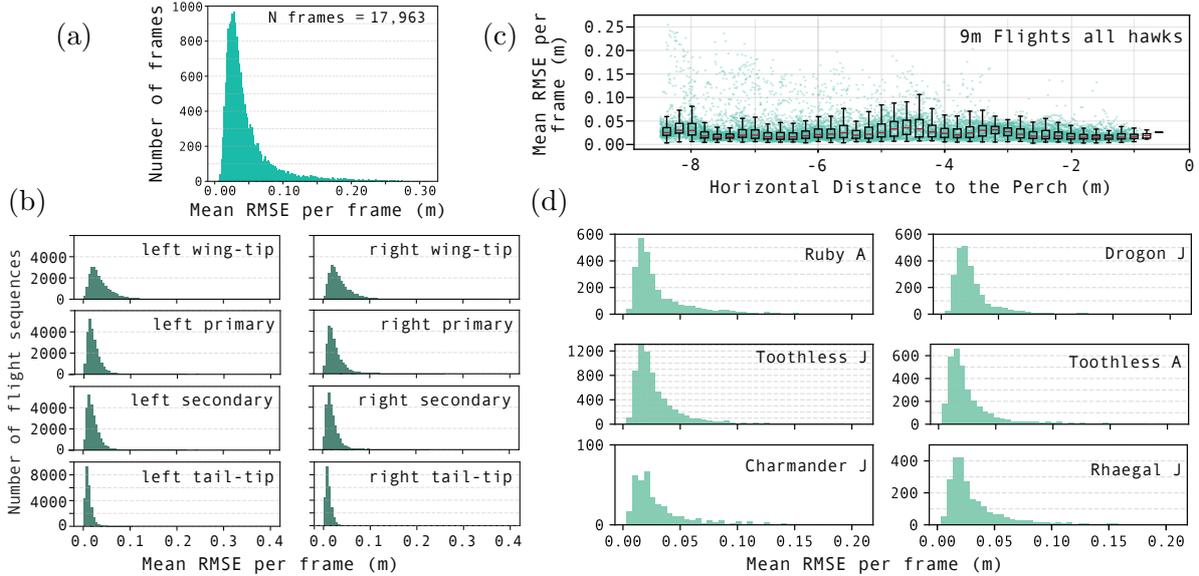
\begin{figure}[t]
  \centering
  \begin{overpic}[width=0.95\textwidth]{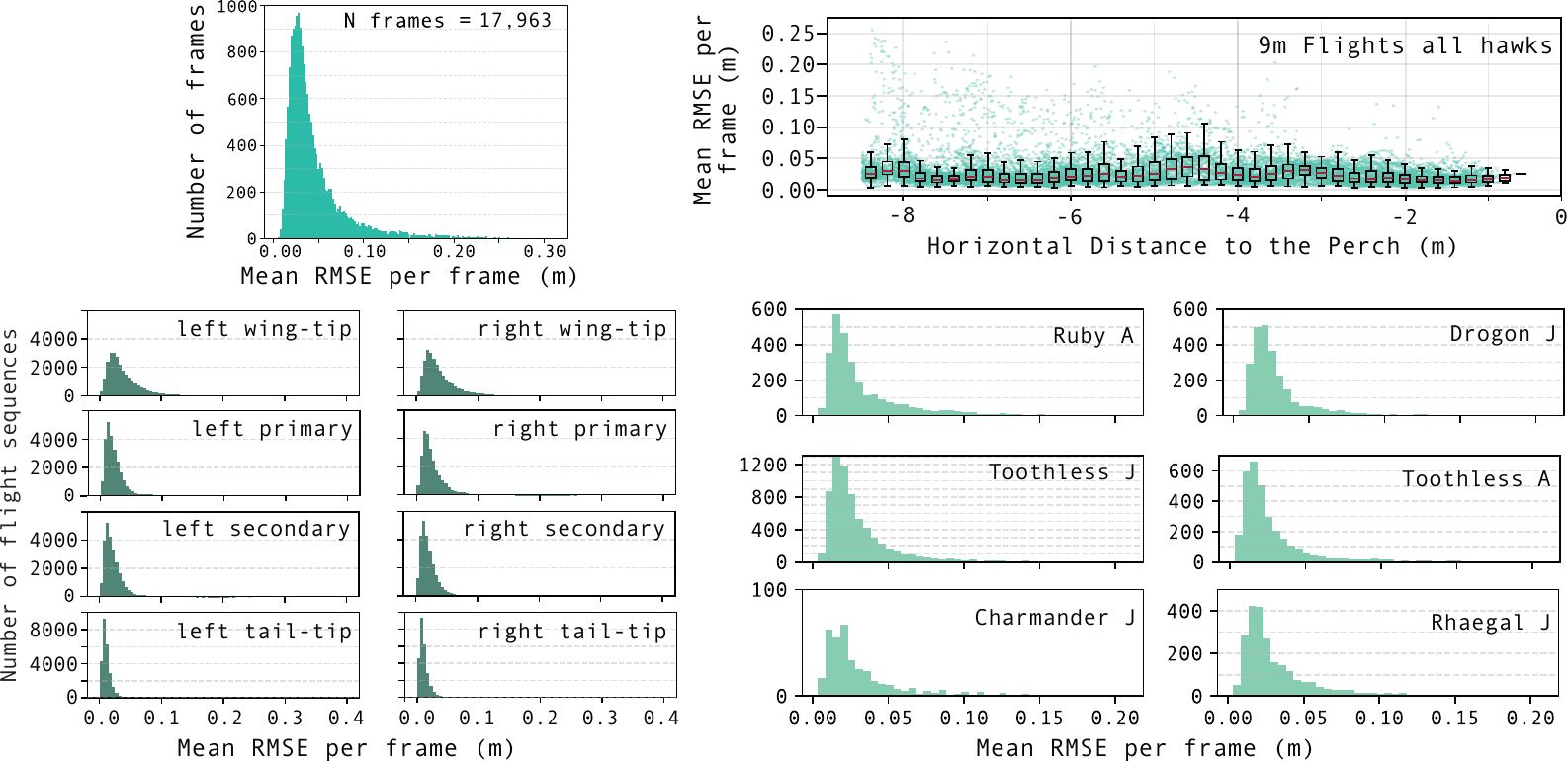}
    \put(4,45){(a)}
    \put(0,31){(b)}
    \put(40,45){(c)}
    \put(44,31){(d)}
    \end{overpic}  
  \caption{Reconstruction error calculated from every DMD run for each individual flight (rather than averaged flights). DMD was run on every flapping flight sequence from the dataset detailed in \cite{France2025}, including multiple perch-perch distances. Root mean squared error (RSME) calculated between the original marker position and the DMD fit. Histograms pool all experimental conditions and show (a) the mean RMSE across markers per frame, pooled from all individuals; (b) the RMSE per marker per frame, pooled from all individuals; (c) the mean RMSE across markers per frame and boxplots binned by horizontal distance to the perch (0.2m), pooling all individuals and using all 9m flights with and without an obstacle; d) the mean RMSE per frame across markers and separated by individual, juvenile and adult as indicated. The wingspan is about a meter, indicating the errors for DMD fits to individual flights are reasonably small.}
  \label{fig:rmse}
\end{figure}

The results shown in Figures \ref{fig:DMDSummary}, \ref{fig:individuals}, \ref{fig:turning} were from binned averages of marker positions over time, using 0.005s bins of 9m flights. For validation, we also fit DMD to every individual flight sequence that passed quality criteria ($N=413$ flight sequences; $n=18,676$ frames) and show the resultant root mean squared errors (RMSE) from reconstruction (Figure \ref{fig:rmse}). This included all seven individual hawks, perch-perch distances of 5, 7, 9, 12m, and the 9m flights either included or excluded an obstacle. For each DMD fit we truncated 6 modes, or three conjugate pairs. We calculated the average RMSE across markers per frame (Figure \ref{fig:rmse}; top left) and find reconstruction error is low from the 6 mode DMD fit. From the average RMSE per frame by marker (bottom left), the wing-tips have the largest absolute RMSE, as they show the largest motion relative to the center of mass. We also fit DMD to the total flight with 6 modes, i.e. including flapping and gliding flight, and show the binned average RMSE by horizontal distance to the perch, showing no significant error increase during the different flight behaviors. Finally we calculate the total RMSE per frame for each of the individual hawks sequence to also confirm no significant error bias. 

\subsection{Broader Connections to Locomotion (Walking) Models}

There are a number of critical connections between bird flight and human walking:  both exhibit periodic motion with two dominant frequencies that are parametrically related, i.e. at approximately $\omega$ and $2\omega$.  Specifically, the earliest models of human walking effectively modeled walking motion as an inverted pendulum system during the stance phase, where the supporting leg acts as a rigid strut while the body's center of mass vaults over it in an arc-like trajectory~\cite{ye2018understand,campanelli2023simple}. This classical model, originally developed by Alexander and Pedley~\cite{alexander1977mechanics} and later refined by Cavagna and Kaneko~\cite{cavagna1977mechanical}, captures fundamental aspects of walking mechanics including the velocity of the center of mass and the exchange between gravitational potential and kinetic energy. However, the connection to a parametrically driven pendulum dynamics becomes particularly relevant when considering how humans restore mechanical energy lost during heel-strike collisions and maintain stable, efficient gait patterns. Research on gait-posture coordination has revealed parametric coupling mechanisms where postural oscillations influence gait dynamics through modulation of effective leg stiffness~\cite{kay2001coupling}, suggesting that human locomotion involves sophisticated parametric excitation strategies beyond simple inverted pendulum mechanics.

The parametric excitation framework offers deeper insights into dynamic walking control and energy efficiency. In parametric excitation walking, mechanical energy dissipated at heel strike is restored through periodic modulation of system parameters—specifically by extending and contracting the legs in coordination with the natural oscillation frequency~\cite{asano2008energy,harata2012efficient}. This mechanism, analogous to pumping the legs while on a swing, enables sustainable gait generation on level ground without requiring continuous energy input from hip actuators. Studies have demonstrated that telescopic leg actuation controlled according to parametric excitation principles can generate high-speed dynamic bipedal gaits with significantly improved energy efficiency~\cite{harata2008parametric,asano2005parametric}. The parametric excitation perspective reveals that human walking likely exploits similar principles through coordinated knee flexion-extension and ankle actuation, where the vertical motion of the center of mass is actively modulated to restore energy and maintain stable limit cycles in the gait dynamics. Similarly, bird flight is driven by a parametric interaction where the center of mass is periodically modulated in flight.

\subsection{Central Pattern Generators}

The dominant oscillation in flight are also suggestive of the existence of an underlying central pattern generator (CPG)~\cite{ijspeert2008central,marder2001central,yuste2005cortex,mackay2002central,pinto2006central}.  CPGs are neural circuits capable of producing rhythmic motor outputs without requiring patterned sensory or descending brain inputs, and they have been documented across diverse species from invertebrates to mammals. The foundational evidence emerged from Wilson's 1961~\cite{Wilson1961} demonstration that isolated locust nervous systems could generate flight-like rhythmic activity, followed by work showing that the mammalian spinal cord can produce basic stepping patterns independently of cortical commands. In humans, spinal cord stimulation studies in individuals with complete spinal cord injuries have demonstrated that lumbar networks can generate rhythmic, locomotor-like movements when isolated from brain control. CPG control is characterized by temperature-dependent frequency scaling and can operate at multiple nested time scales to coordinate complex behaviors from simpler subroutines. The shared modal structure we observe in hawk flight across individuals, despite high individual variation in flight technique, raises important questions about whether similar CPG-like circuits underlie avian flight control. The parametric frequency doubling and the gradual attenuating-amplifying interplay between modes suggest an optimized control architecture that balances energy efficiency with maneuverability. Future work employing windowed DMD methods could resolve how these modes adapt during rapid transitions and turning, while comparative studies across species could reveal whether the three-mode structure represents a universal feature of powered vertebrate flight or an adaptation specific to hawks and similarly sized raptors. The DMD approach demonstrated here provides a foundation for understanding flight control as the coordinated activation of low-dimensional motor primitives, bridging the gap between high-dimensional morphing kinematics and the neural computations that generate them.

%\subsection{Base model of flight}

%\subsection{Gliding mode}

%Underlies the flapping dynamics and takes over as the main dynamics as the flapping turns off.

%\subsection{Turning modes}

%- Mirrored modes for the turning
%- Describe the three layers of simultaneous movement.
%- DMD appears to be able to decompose multiple competing dynamical constraints on the birds movement during even more complex maneuvers such as turning. From this we can recover modes describing these underlying individual components <- maybe a discussion point.

\subsection{Outlook}

This study demonstrates that the complex flight dynamics of Harris' hawks can be accurately reconstructed using a minimal set of interpretable dynamic modes derived from motion capture data. Through Dynamic Mode Decomposition, we identified three fundamental modes that characterize flapping flight: a primary wingbeat oscillation at the fundamental frequency (approximately 4.5 Hz), a secondary modulation at nearly twice this frequency (approximately 8.8 Hz), and a non-oscillatory mode capturing the gradual transition toward gliding posture. This low-dimensional representation achieves reconstruction errors within 1.19\% of maximum wingspan while providing mechanistic insight into how hawks coordinate wing-tail morphing during natural flight. Critically, these modes are shared across individuals despite substantial variation in flight style, suggesting they represent fundamental biomechanical constraints or control strategies common to hawk flight.

The parametric relationship between the first two oscillatory modes—with frequencies at approximately $\omega$ and $2\omega$—reveals a striking parallel to parametric excitation mechanisms observed in human walking. In bipedal locomotion, parametric coupling between dominant frequencies enables energy-efficient gait generation through coordinated modulation of system parameters. The doubled-frequency mode in hawk flight exhibits characteristics consistent with an active control mechanism rather than passive aeroelasticity: it is phase-locked to the wingbeat cycle, highly symmetric across left and right wings, and localized to the distal wing where fine-scale aerodynamic control is most effective. We propose that this mode functions analogously to distal forearm muscle activation patterns observed in pigeons, where alternating pronator and supinator activity within each wingbeat produces torsional adjustments that modulate lift and thrust production. The 5\% reduction in integrated wingtip velocity upon removal of this mode supports its functional aerodynamic significance despite its relatively small amplitude.  

These results are also useful to those designing morphing wing UAVs. Engineering-informed biology experiments could test the aerodynamic effect of these dynamic modes directly, using morphing wing designs that mimic the degrees of freedom of birds, and perhaps inform new coordinated control of morphing. For bio-informed engineering, it can be suboptimal to directly copy wingbeat flapping and forward power is more efficient instead with rotatory blades. DMD effectively separates out the flapping wingbeat from other morphing dynamics used to generate maneuvering flight. 

Importantly, the DMD analysis framework presented offers substantial advantages over traditional bottom-up flight models. Unlike kinematic prescriptions or quasi-steady aerodynamic assumptions, our data-driven approach captures the temporal dynamics and multi-objective nature of real flight without imposing unrealistic constraints. The emergence of the gliding mode as a continuously varying contribution rather than a discrete behavioral switch validates previous observations that flight behaviors exist on a continuum. Furthermore, the model's ability to decompose turning maneuvers into mirrored banking modes and asymmetric perturbations demonstrates that DMD can disentangle multiple simultaneous constraints even during complex multi-objective flight. This interpretability, combined with the model's generative capacity for forward prediction, suggests that DMD-derived modes may represent fundamental building blocks of avian flight control—potentially reflecting an underlying neural pattern generator that combines stereotyped motor primitives to achieve diverse flight behaviors.

The shared modal structure across individuals, despite high individual variation in flight technique, raises important questions about the evolutionary and biomechanical origins of these dynamic patterns. The parametric frequency doubling and the gradual attenuating-amplifying interplay between modes suggest an optimized control architecture that balances energy efficiency with maneuverability. Future work employing windowed DMD methods could resolve how these modes adapt during rapid transitions and turning, while comparative studies across species could reveal whether the three-mode structure represents a universal feature of powered vertebrate flight or an adaptation specific to hawks and similarly sized raptors. The DMD approach demonstrated here provides a foundation for understanding flight control as the coordinated activation of low-dimensional motor primitives, bridging the gap between high-dimensional morphing kinematics and the neural computations that generate them.

\section*{Acknowledgements}

This work was supported in part by the US National Science Foundation (NSF) AI Institute for Dynamical Systems (dynamicsai.org), grant 2112085. JNK further acknowledges support from the Air Force Office of Scientific Research  (FA9550-24-1-0141). K.L. was funded by the Austrian Science Fund (FWF) [doi.org/10.55776/ESP214]. This project has received funding from the European Research Council (ERC) under the European Union’s Horizon 2020 Research and Innovation Programme (grant agreement no. 682501). LAF was supported by Schmidt Sciences, LLC.  We thank Steve Portugal and Ben Lambert for their useful comments on the draft manuscript. We thank Kirsty Yeomans for her illustration in Figure 1a. 

\section*{Code and Data}

A public repository containing the analysis code and accompanying Jupyter notebooks demonstrating the analysis in this study is available at \href{https://github.com/LydiaFrance/BirdDMD}{https://github.com/LydiaFrance/BirdDMD}, with rendered documentation and visualisations at \href{https://lydiafrance.github.io/BirdDMD/}{https://lydiafrance.github.io/BirdDMD/}. These include animations of flight kinematics, DMD modes, reconstructions, and generative bird flight. The repository provides executable workflows corresponding to the methods and figures presented in the manuscript, enabling readers to inspect and run the computational analyses.

\bibliographystyle{unsrtsiam}
\bibliography{refs}

@article{robust_dmd,
author = {Askham, Travis and Zheng, Peng and Aravkin, Aleksandr and Kutz, J. Nathan},
title = {Robust and Scalable Methods for the Dynamic Mode Decomposition},
journal = {SIAM Journal on Applied Dynamical Systems},
volume = {21},
number = {1},
pages = {60-79},
year = {2022},
doi = {10.1137/21M1417405},
URL = {https://doi.org/10.1137/21M1417405},
eprint = {https://doi.org/10.1137/21M1417405},
}

@article{Wilson1961,
  author = {Wilson, Donald M.},
  title = {The central nervous control of flight in a locust},
  journal = {Journal of Experimental Biology},
  year = {1961},
  volume = {38},
  number = {2},
  pages = {471--490}
}

@article{ijspeert2008central,
  title={Central pattern generators for locomotion control in animals and robots: a review},
  author={Ijspeert, Auke Jan},
  journal={Neural networks},
  volume={21},
  number={4},
  pages={642--653},
  year={2008},
  publisher={Elsevier}
}

@article{marder2001central,
  title={Central pattern generators and the control of rhythmic movements},
  author={Marder, Eve and Bucher, Dirk},
  journal={Current biology},
  volume={11},
  number={23},
  pages={R986--R996},
  year={2001},
  publisher={Elsevier}
}

@article{yuste2005cortex,
  title={The cortex as a central pattern generator},
  author={Yuste, Rafael and MacLean, Jason N and Smith, Jeffrey and Lansner, Anders},
  journal={Nature Reviews Neuroscience},
  volume={6},
  number={6},
  pages={477--483},
  year={2005},
  publisher={Nature Publishing Group UK London}
}

@article{pinto2006central,
  title={Central pattern generators for bipedal locomotion},
  author={Pinto, Carla MA and Golubitsky, Martin},
  journal={Journal of mathematical biology},
  volume={53},
  number={3},
  pages={474--489},
  year={2006},
  publisher={Springer}
}

@article{mackay2002central,
  title={Central pattern generation of locomotion: a review of the evidence},
  author={MacKay-Lyons, Marilyn},
  journal={Physical therapy},
  volume={82},
  number={1},
  pages={69--83},
  year={2002},
  publisher={Oxford University Press}
}

@article{optdmd_c,
title = {Constrained optimized dynamic mode decomposition with control for physically stable systems with exogeneous inputs},
journal = {Journal of Computational Physics},
volume = {496},
pages = {112604},
year = {2024},
issn = {0021-9991},
doi = {https://doi.org/10.1016/j.jcp.2023.112604},
url = {https://www.sciencedirect.com/science/article/pii/S002199912300699X},
author = {Jacob Rains and Yi Wang and Alec House and Andrew L. Kaminsky and Nathan A. Tison and Vamshi M. Korivi},
}

@inproceedings{ye2018understand,
  title={Understand human walking through a 2D inverted pendulum model},
  author={Ye, Linqi and Chen, Xuechao},
  booktitle={2018 IEEE-RAS 18th International Conference on Humanoid Robots (Humanoids)},
  pages={340--345},
  year={2018},
  organization={IEEE}
}

@article{campanelli2023simple,
  title={A simple model of human walking},
  author={Campanelli, Leonardo},
  journal={Journal of Medical Science},
  volume={92},
  number={1},
  pages={e817--e817},
  year={2023}
}

@article{alexander1977mechanics,
  title={Mechanics and scaling of terrestrial locomotion},
  author={Alexander, R McNeill and Pedley, TJ},
  journal={Scale effects in animal locomotion},
  pages={93--110},
  year={1977},
  publisher={Academic Press London}
}

@article{cavagna1977mechanical,
  title={Mechanical work and efficiency in level walking and running},
  author={Cavagna, GA and Kaneko, MJTJop},
  journal={The Journal of physiology},
  volume={268},
  number={2},
  pages={467--481},
  year={1977},
  publisher={Wiley Online Library}
}

@article{kay2001coupling,
  title={Coupling of posture and gait: mode locking and parametric excitation},
  author={Kay, Bruce A and Warren Jr, William H},
  journal={Biological cybernetics},
  volume={85},
  number={2},
  pages={89--106},
  year={2001},
  publisher={Springer}
}

@article{asano2008energy,
  title={Energy-efficient and high-speed dynamic biped locomotion based on principle of parametric excitation},
  author={Asano, Fumihiko and Luo, Zhi-Wei},
  journal={IEEE Transactions on Robotics},
  volume={24},
  number={6},
  pages={1289--1301},
  year={2008},
  publisher={IEEE}
}

@article{harata2012efficient,
  title={Efficient parametric excitation walking with delayed feedback control},
  author={Harata, Yuji and Asano, Fumihiko and Taji, Kouichi and Uno, Yoji},
  journal={Nonlinear Dynamics},
  volume={67},
  number={2},
  pages={1327--1335},
  year={2012},
  publisher={Springer}
}

@inproceedings{harata2008parametric,
  title={Parametric excitation based gait generation for ornithoid walking},
  author={Harata, Yuji and Asano, Fumihiko and Taji, Kouichi and Uno, Yoji},
  booktitle={2008 IEEE/RSJ International Conference on Intelligent Robots and Systems},
  pages={2940--2945},
  year={2008},
  organization={IEEE}
}

@inproceedings{asano2005parametric,
  title={Parametric excitation mechanisms for dynamic bipedal walking},
  author={Asano, Fumihiko and Luo, Zhi-Wei and Hyon, S},
  booktitle={Proceedings of the 2005 IEEE International Conference on Robotics and Automation},
  pages={609--615},
  year={2005},
  organization={IEEE}
}

@book{dmd_book,
author = {Kutz, J. Nathan and Brunton, Steven L. and Brunton, Bingni W. and Proctor, Joshua L.},
title = {Dynamic Mode Decomposition: Data-Driven Modeling of Complex Systems},
publisher = {Society for Industrial and Applied Mathematics},
year = {2016},
doi = {10.1137/1.9781611974508},
address = {Philadelphia, PA},
URL = {https://epubs.siam.org/doi/abs/10.1137/1.9781611974508},
eprint = {https://epubs.siam.org/doi/pdf/10.1137/1.9781611974508},
note = {\href{https://doi.org/10.1137/1.9781611974508}{https://doi.org/10.1137/1.9781611974508}}
}

@book{kutz2013data,
  title={Data-driven modeling \& scientific computation: methods for complex systems \& big data},
  author={Kutz, Jose Nathan},
  year={2013},
  publisher={OUP Oxford}
}

@book{data_book,
author = {Brunton, Steven L. and Kutz, J. Nathan},
title = {Data-Driven Science and Engineering: Machine Learning, Dynamical Systems, and Control},
publisher = {Cambridge University Press},
year = {2022},
doi = {10.1017/9781009089517},
URL = {https://doi.org/10.1017/9781009089517},
note = {\href{https://doi.org/10.1017/9781009089517}{https://doi.org/10.1017/9781009089517}}
}

@InProceedings{schmid_2008,
author = {Peter J. Schmid and Joern Sesterhenn},
title = {Dynamic Mode Decomposition of Numerical and Experimental Data},
booktitle = {61st Annual Meeting of the APS Division of Fluid Dynamics},
organization = {American Physical Society},
year = {2008},
note = {\href{http://meetings.aps.org/link/BAPS.2008.DFD.MR.7}{http://meetings.aps.org/link/BAPS.2008.DFD.MR.7}}
}

@article{schmid_2010,
author = {Peter J. Schmid},
title = {Dynamic mode decomposition of numerical and experimental data},
volume = {656},
pages = {5-28},
journal = {Journal of Fluid Mechanics}, 
publisher = {Cambridge University Press}, 
year = {2010},
doi = {10.1017/S0022112010001217},
url = {https://doi.org/10.1017/S0022112010001217},
note = {\href{https://doi.org/10.1017/S0022112010001217}{https://doi.org/10.1017/S0022112010001217}}
}

@inproceedings{schmid_2009,
TITLE = {Dynamic mode decomposition of experimental data},
AUTHOR = {Schmid, Peter J.},
URL = {https://polytechnique.hal.science/hal-01053394},
BOOKTITLE = {{8th International Symposium on Particle Image Velocimetry (PIV09)}},
ADDRESS = {Melbourne, Australia},
YEAR = {2009},
HAL_ID = {hal-01053394},
HAL_VERSION = {v1},
note = {\href{https://polytechnique.hal.science/hal-01053394}{https://polytechnique.hal.science/hal-01053394}}
}

@article{proctor_2015,
title = {Discovering dynamic patterns from infectious disease data using dynamic mode decomposition},
author = {Proctor, Joshua L. and Eckhoff, Philip A.},
journal = {International Health},
volume = {7},
number = {2},
pages = {139-145},
year = {2015},
doi = {10.1093/inthealth/ihv009},
URL = {https://doi.org/10.1093/inthealth/ihv009},
eprint = {https://doi.org/10.1093/inthealth/ihv009},
note = {\href{https://doi.org/10.1093/inthealth/ihv009}{https://doi.org/10.1093/inthealth/ihv009}}
}

@article{brunton_2016,
title = {Extracting spatial–temporal coherent patterns in large-scale neural recordings using dynamic mode decomposition},
author = {Bingni W. Brunton and Lise A. Johnson and Jeffrey G. Ojemann and J. Nathan Kutz},
journal = {Journal of Neuroscience Methods},
volume = {258},
pages = {1-15},
year = {2016},
doi = {10.1016/j.jneumeth.2015.10.010},
URL = {https://doi.org/10.1016/j.jneumeth.2015.10.010},
eprint = {https://doi.org/10.1016/j.jneumeth.2015.10.010},
note = {\href{https://doi.org/10.1016/j.jneumeth.2015.10.010}{https://doi.org/10.1016/j.jneumeth.2015.10.010}}
}

@article{alfatlawi_2020,
title = {An incremental approach to online dynamic mode decomposition for time-varying systems with applications to {EEG} data modeling},
author = {Mustaffa Alfatlawi and Vaibhav Srivastava},
journal = {Journal of Computational Dynamics},
volume = {7},
number = {2},
pages = {209-241},
year = {2020},
doi = {10.3934/jcd.2020009},
URL = {https://doi.org/10.3934/jcd.2020009},
eprint = {https://doi.org/10.3934/jcd.2020009},
note = {\href{https://doi.org/10.3934/jcd.2020009}{https://doi.org/10.3934/jcd.2020009}}
}

@article{mann_2016,
author = {Jordan Mann and J. Nathan Kutz},
title = {Dynamic mode decomposition for financial trading strategies},
journal = {Quantitative Finance},
volume = {16},
pages = {1643-1655},
year  = {2016},
publisher = {Routledge},
doi = {10.1080/14697688.2016.1170194},
URL = {https://doi.org/10.1080/14697688.2016.1170194},
note = {\href{https://doi.org/10.1080/14697688.2016.1170194}{https://doi.org/10.1080/14697688.2016.1170194}}
}

@article{taylor_2018,
title = {Dynamic mode decomposition for plasma diagnostics and validation},
author = {Roy Taylor and J. Nathan Kutz and Kyle Morgan and Brian A. Nelson},
journal = {Review of Scientific Instruments},
volume = {89},
number = {5},
year = {2018},
pages = {053501},
doi = {10.1063/1.5027419},
URL = {https://doi.org/10.1063/1.5027419},
eprint = {https://doi.org/10.1063/1.5027419},
note = {\href{https://doi.org/10.1063/1.5027419}{https://doi.org/10.1063/1.5027419}}
}

@article{kaptanoglu_2020,
title = {Characterizing magnetized plasmas with dynamic mode decomposition},
author = {A. A. Kaptanoglu and K. D. Morgan and C. J. Hansen and S. L. Brunton},
journal = {Physics of Plasmas},
volume = {27},
year = {2020},
pages = {032108},
doi = {10.1063/1.5138932},
URL = {https://doi.org/10.1063/1.5138932},
eprint = {https://doi.org/10.1063/1.5138932},
note = {\href{https://doi.org/10.1063/1.5138932}{https://doi.org/10.1063/1.5138932}}
}

@article{berger_2015,
author = {Erik Berger and Mark Sastuba and David Vogt and Bernhard Jung and Heni Ben Amor},
title = {Estimation of perturbations in robotic behavior using dynamic mode decomposition},
journal = {Advanced Robotics},
volume = {29},
pages = {331-343},
year  = {2015},
publisher = {Taylor & Francis},
doi = {10.1080/01691864.2014.981292},
URL = {https://doi.org/10.1080/01691864.2014.981292},
note = {\href{https://doi.org/10.1080/01691864.2014.981292}{https://doi.org/10.1080/01691864.2014.981292}}
}

@article{abraham_2019,
author={Abraham, Ian and Murphey, Todd D.},
journal={IEEE Transactions on Robotics}, 
title={Active Learning of Dynamics for Data-Driven Control Using {K}oopman Operators}, 
year={2019},
volume={35},
number={5},
pages={1071-1083},
doi={10.1109/TRO.2019.2923880},
note={\href{https://doi.org/10.1109/TRO.2019.2923880}{https://doi.org/10.1109/TRO.2019.2923880}}
}

@inproceedings{bruder_2019,
author = {Daniel Bruder and Brent Gillespie and C. David Remy and Ram Vasudevan},
title = {Modeling and Control of Soft Robots Using the {K}oopman Operator and Model Predictive Control},
booktitle = {Robotics: Science and Systems XV},
year = {2019},
note = {\href{https://doi.org/10.15607/RSS.2019.XV.060}{https://doi.org/10.15607/RSS.2019.XV.060}}
}

@article{rowley_2009,
author = {Clarence W. Rowley and Igor Mezi\'{c} and Shervin Bagheri and Philipp Schlatter and Dan S. Henningson},
title = {Spectral Analysis of Nonlinear Flows},
volume = {641},
journal = {Journal of Fluid Mechanics},
publisher = {Cambridge University Press},
year = {2009},
pages = {115-127},
doi = {10.1017/S0022112009992059},
note = {\href{https://doi.org/10.1017/S0022112009992059}{https://doi.org/10.1017/S0022112009992059}}
}

@article{tu_2014,
author = {Jonathan H. Tu and Clarence W. Rowley and Dirk M. Luchtenburg and Steven L. Brunton and J. N. Kutz},
title = {On Dynamic Mode Decomposition: Theory and Applications},
year = {2014},
pages = {391-421},
volume = {1},
journal = {Journal of Computational Dynamics},
doi = {10.3934/jcd.2014.1.391},
note = {\href{https://doi.org/10.3934/jcd.2014.1.391}{https://doi.org/10.3934/jcd.2014.1.391}}
}

@article{modern_koopman,
author = {Brunton, Steven L. and Budi\v{s}i\'{c}, Marko and Kaiser, Eurika and Kutz, J. Nathan},
title = {Modern {K}oopman Theory for Dynamical Systems},
journal = {SIAM Review},
volume = {64},
number = {2},
pages = {229-340},
year = {2022},
doi = {10.1137/21M1401243},
URL = {https://doi.org/10.1137/21M1401243},
eprint = {https://doi.org/10.1137/21M1401243},
note = {\href{https://doi.org/10.1137/21M1401243}{https://doi.org/10.1137/21M1401243}}
}

@article{pydmd,
author = {Nicola Demo and Marco Tezzele and Gianluigi Rozza},
title = {{P}y{DMD}: {P}ython Dynamic Mode Decomposition},
journal = {Journal of Open Source Software},
year = {2018},
publisher = {The Open Journal},
volume = {3},
number = {22},
pages = {530},
doi = {10.21105/joss.00530},
url = {https://doi.org/10.21105/joss.00530},
note = {\href{https://doi.org/10.21105/joss.00530}{https://doi.org/10.21105/joss.00530}}
}

@article{optdmd,
author = {Askham, Travis and Kutz, J. Nathan},
title = {Variable Projection Methods for an Optimized Dynamic Mode Decomposition},
journal = {SIAM Journal on Applied Dynamical Systems},
volume = {17},
number = {1},
pages = {380-416},
year = {2018},
doi = {10.1137/M1124176},
URL = {https://doi.org/10.1137/M1124176},
eprint = {https://doi.org/10.1137/M1124176},
note = {\href{https://doi.org/10.1137/M1124176}{https://doi.org/10.1137/M1124176}}
}

@article{bopdmd,
title = {Bagging, optimized dynamic mode decomposition for robust, stable forecasting with spatial and temporal uncertainty quantification},
author = {Sashidhar, Diya and Kutz, J. Nathan},
journal = {Proceedings of the Royal Society A},
volume = {380},
number = {2229},
year = {2022},
pages = {20210199},
doi = {10.1098/rsta.2021.0199},
URL = {https://doi.org/10.1098/rsta.2021.0199},
eprint = {https://doi.org/10.1098/rsta.2021.0199},
note = {\href{https://doi.org/10.1098/rsta.2021.0199}{https://doi.org/10.1098/rsta.2021.0199}}
}

@article{schmid2022dynamic,
author = {Schmid, Peter J},
journal = {Annual Review of Fluid Mechanics},
pages = {225--254},
title = {Dynamic mode decomposition and its variants},
volume = {54},
year = {2022},
note = {\href{https://doi.org/10.1146/annurev-fluid-030121-015835}{https://doi.org/10.1146/annurev-fluid-030121-015835}}}

@article{Noack2016jfm,
author = {Noack, Bernd R and Stankiewicz, Witold and Morzy\'{n}ski, Marek and Schmid, Peter J},
journal = {Journal of Fluid Mechanics},
pages = {843--872},
title = {Recursive dynamic mode decomposition of transient and post-transient wake flows},
volume = {809},
year = {2016},
note = {\href{https://doi.org/10.1017/jfm.2016.678}{https://doi.org/10.1017/jfm.2016.678}}
}

@article {France2025,
	author = {France, Lydia A. and Shelton, James and KleinHeerenbrink, Marco and Brighton, Caroline and Taylor, Graham K.},
	title = {Signatures of Motion: Decomposition of Adaptive Morphing Flight in {Harris}{\textquoteright} {Hawks}},
	elocation-id = {2025.06.03.657611},
	year = {2025},
	doi = {10.1101/2025.06.03.657611},
	publisher = {Cold Spring Harbor Laboratory},
	URL = {https://www.biorxiv.org/content/early/2025/06/07/2025.06.03.657611},
	eprint = {https://www.biorxiv.org/content/early/2025/06/07/2025.06.03.657611.full.pdf},
	journal = {bioRxiv}, 
    note = {\href{https://doi.org/10.1101/2025.06.03.657611}{https://doi.org/10.1101/2025.06.03.657611}}
}

@article{chin2019b,
  title = {Birds Repurpose the Role of Drag and Lift to Take off and Land},
  author = {Chin, Diana D. and Lentink, David},
  date = {2019-11-25},
  year = {2019},
  journal = {Nature Communications},
  shortjournal = {Nat Commun},
  volume = {10},
  number = {1},
  pages = {5354},
  publisher = {Nature Publishing Group},
  issn = {2041-1723},
  doi = {10.1038/s41467-019-13347-3},
  url = {https://www.nature.com/articles/s41467-019-13347-3},
  urldate = {2025-12-16},
  langid = {english}
}

@article{altshuler2015a,
  title = {The Biophysics of Bird Flight: Functional Relationships Integrate Aerodynamics, Morphology, Kinematics, Muscles, and Sensors},
  shorttitle = {The Biophysics of Bird Flight},
  author = {Altshuler, Douglas L. and Bahlman, Joseph W. and Dakin, Roslyn and Gaede, Andrea H. and Goller, Benjamin and Lentink, David and Segre, Paolo S. and Skandalis, Dimitri A.},
  date = {2015-12},
  year = {2015},
  journal = {Canadian Journal of Zoology},
  shortjournal = {Can. J. Zool.},
  volume = {93},
  number = {12},
  pages = {961--975},
  issn = {0008-4301, 1480-3283},
  doi = {10.1139/cjz-2015-0103},
  url = {http://www.nrcresearchpress.com/doi/10.1139/cjz-2015-0103},
  urldate = {2024-05-26},
  langid = {english}
}

@article{chin2016a,
  title = {Flapping Wing Aerodynamics: From Insects to Vertebrates},
  shorttitle = {Flapping Wing Aerodynamics},
  author = {Chin, Diana D. and Lentink, David},
  date = {2016-04-01},
  year = {2016},
  journal = {Journal of Experimental Biology},
  shortjournal = {Journal of Experimental Biology},
  volume = {219},
  number = {7},
  pages = {920--932},
  issn = {0022-0949},
  doi = {10.1242/jeb.042317},
  url = {https://doi.org/10.1242/jeb.042317},
  urldate = {2024-05-26}
}

@article{dakin_individual_2020,
	title = {Individual variation and the biomechanics of maneuvering flight in hummingbirds},
	volume = {223},
	issn = {0022-0949},
	url = {https://doi.org/10.1242/jeb.161828},
	doi = {10.1242/jeb.161828},
	number = {20},
	journal = {Journal of Experimental Biology},
	shortjournal = {Journal of Experimental Biology},
	author = {Dakin, R. and Segre, P. S. and Altshuler, D. L.},
	urldate = {2025-05-30},
	date = {2020-10-27},
    year = {2020}
}

@article{ellington1984,
  title = {The Aerodynamics of Hovering Insect Flight. {{I}}. {{The}} Quasi-Steady Analysis},
  author = {Ellington, Charles Porter},
  date = {1984-02-24},
  year = {1984},
  journal = {Philosophical Transactions of the Royal Society of London. B, Biological Sciences},
  shortjournal = {Philos Trans R Soc Lond B Biol Sci},
  volume = {305},
  number = {1122},
  pages = {1--15},
  issn = {0080-4622},
  doi = {10.1098/rstb.1984.0049},
  url = {https://doi.org/10.1098/rstb.1984.0049},
  urldate = {2025-12-16}
}

@article{ellington1996,
  title = {Leading-Edge Vortices in Insect Flight},
  author = {Ellington, Charles P. and van den Berg, Coen and Willmott, Alexander P. and Thomas, Adrian L. R.},
  date = {1996-12},
  year = {1996},
  journal = {Nature},
  volume = {384},
  number = {6610},
  pages = {626--630},
  publisher = {Nature Publishing Group},
  issn = {1476-4687},
  doi = {10.1038/384626a0},
  url = {https://www.nature.com/articles/384626a0},
  urldate = {2025-12-16},
  langid = {english}
}

@article{harvey2022d,
  title = {A Review of Avian-Inspired Morphing for {{UAV}} Flight Control},
  author = {Harvey, Christina and Gamble, Lawren L. and Bolander, Christian R. and Hunsaker, Douglas F. and Joo, James J. and Inman, Daniel J.},
  year = {2022},
  date = {2022-07-01},
  journal = {Progress in Aerospace Sciences},
  shortjournal = {Progress in Aerospace Sciences},
  volume = {132},
  pages = {100825},
  issn = {0376-0421},
  doi = {10.1016/j.paerosci.2022.100825},
  url = {https://www.sciencedirect.com/science/article/pii/S0376042122000173},
  urldate = {2025-12-16}
}

@article{perching_2022,
	title = {Optimization of avian perching manoeuvres},
	volume = {607},
	copyright = {2022 The Author(s)},
	issn = {1476-4687},
	url = {https://www.nature.com/articles/s41586-022-04861-4},
	doi = {10.1038/s41586-022-04861-4},
	language = {en},
	number = {7917},
	urldate = {2024-07-29},
	journal = {Nature},
	publisher = {Nature Publishing Group},
	author = {KleinHeerenbrink, Marco and France, Lydia A. and Brighton, Caroline H. and Taylor, Graham K.},
	month = jul,
	year = {2022},
	pages = {91--96},
    note = {\href{https://doi.org/10.1038/s41586-022-04861-4}{https://doi.org/10.1038/s41586-022-04861-4}}
}

@article{martinez_groves-raines_steady_2024,
	title = {Steady as they hover: kinematics of kestrel wing and tail morphing during hovering flights},
	volume = {227},
	issn = {0022-0949},
	url = {https://doi.org/10.1242/jeb.247305},
	doi = {10.1242/jeb.247305},
	number = {15},
	journal = {Journal of Experimental Biology},
	author = {Martinez Groves-Raines, Mario and Yi, George and Penn, Matthew and Watkins, Simon and Windsor, Shane and Mohamed, Abdulghani},
	month = aug,
	year = {2024},
	url = {10.1242/jeb.247305},
	pages = {},
}

@article{pennycuick1968a,
  title = {Power {{Requirements}} for {{Horizontal Flight}} in the {{Pigeon Columba Livia}}},
  author = {Pennycuick, C. J.},
  date = {1968-12-01},
  year = {1968},
  journal = {Journal of Experimental Biology},
  shortjournal = {J Exp Biol},
  volume = {49},
  number = {3},
  pages = {527--555},
  issn = {0022-0949},
  doi = {10.1242/jeb.49.3.527},
  url = {https://doi.org/10.1242/jeb.49.3.527},
  urldate = {2025-12-16}
}

@book{pennycuick2008,
  title = {Modelling the {{Flying Bird}}},
  author = {Pennycuick, C. J.},
  year = {2008},
  publisher = {Elsevier Science \& Technology},
  location = {Chantilly, UNITED STATES},
  url = {http://ebookcentral.proquest.com/lib/oxford/detail.action?docID=404801},
  urldate = {2025-12-16},
  isbn = {978-0-08-055781-6}
}

@article{rayner1979,
  title = {A Vortex Theory of Animal Flight. {{Part}} 1. {{The}} Vortex Wake of a Hovering Animal},
  author = {Rayner, J. M. V.},
  date = {1979-04},
  year = {1979},
  journal = {Journal of Fluid Mechanics},
  volume = {91},
  number = {4},
  pages = {697--730},
  issn = {1469-7645, 0022-1120},
  doi = {10.1017/S0022112079000410},
  url = {https://www.cambridge.org/core/journals/journal-of-fluid-mechanics/article/vortex-theory-of-animal-flight-part-1-the-vortex-wake-of-a-hovering-animal/56480F7077534D93939F601F29F7A7A9},
  urldate = {2025-12-16},
  langid = {english}
}

@article{rayner2001,
  title = {Mathematical Modelling of the Avian Flight Power Curve},
  author = {Rayner, Jeremy M. V.},
  year = {2001},
  journal = {Mathematical Methods in the Applied Sciences},
  volume = {24},
  number = {17--18},
  pages = {1485--1514},
  issn = {1099-1476},
  doi = {10.1002/mma.196},
  url = {https://onlinelibrary.wiley.com/doi/abs/10.1002/mma.196},
  urldate = {2025-12-16},
  langid = {english}
}

@article{robertson2012,
  title = {Muscle Function during Takeoff and Landing Flight in the Pigeon ({{Columba}} Livia)},
  author = {Robertson, Angela M. Berg and Biewener, Andrew A.},
  date = {2012-12-01},
  year = 2012,
  journal = {Journal of Experimental Biology},
  shortjournal = {J Exp Biol},
  volume = {215},
  number = {23},
  pages = {4104--4114},
  issn = {0022-0949},
  doi = {10.1242/jeb.075275},
  url = {https://doi.org/10.1242/jeb.075275},
  urldate = {2026-01-20}
}

@article{shen2025,
  title = {Effect of {{Coupled Wing Motion}} on the {{Aerodynamic Performance}} during {{Different Flight Stages}} of {{Pigeon}}},
  author = {Shen, Yishi and Xu, Yi and Huang, Weimin and Shang, Chengrui and Shi, Qing},
  date = {2025-03-11},
  year = 2025,
  journal = {Cyborg and Bionic Systems},
  volume = {6},
  pages = {0200},
  publisher = {American Association for the Advancement of Science},
  doi = {10.34133/cbsystems.0200},
  url = {https://spj.science.org/doi/10.34133/cbsystems.0200},
  urldate = {2025-12-16}
}

@article{song2022a,
  title = {Virtual Manipulation of Tail Postures of a Gliding Barn Owl ( {{Tyto}} Alba ) Demonstrates Drag Minimization When Gliding},
  author = {Song, Jialei and Cheney, Jorn A. and Bomphrey, Richard J. and Usherwood, James R.},
  date = {2022-02-01},
  year = {2022},
  journal = {Journal of the Royal Society Interface},
  volume = {19},
  number = {187},
  doi = {10.1098/rsif.2021.0710}
}

@article{taylor2002,
  title = {Animal {{Flight Dynamics II}}. {{Longitudinal Stability}} in {{Flapping Flight}}},
  author = {Taylor, G. K. and Thomas, A. L. R.},
  date = {2002-02-07},
  year = 2002,
  journal = {Journal of Theoretical Biology},
  shortjournal = {Journal of Theoretical Biology},
  volume = {214},
  number = {3},
  pages = {351--370},
  issn = {0022-5193},
  doi = {10.1006/jtbi.2001.2470},
  url = {https://www.sciencedirect.com/science/article/pii/S0022519301924701},
  urldate = {2025-11-18}
}

@article{tobalske2022b,
  title = {Aerodynamics of Avian Flight},
  author = {Tobalske, Bret W.},
  date = {2022-10-24},
  year = {2022},
  journal = {Current Biology},
  shortjournal = {Current Biology},
  volume = {32},
  number = {20},
  pages = {R1105-R1109},
  issn = {0960-9822},
  doi = {10.1016/j.cub.2022.07.007},
  url = {https://www.sciencedirect.com/science/article/pii/S0960982222011083},
  urldate = {2025-12-16}
}

@article{usherwood2002,
  title = {The Aerodynamics of Revolving Wings {{II}}. {{Propeller}} Force Coefficients from Mayfly to Quail},
  author = {Usherwood, James R. and Ellington, Charles P.},
  year = {2002},
  journal = {Journal of Experimental Biology},
  volume = {205},
  number = {11},
  pages = {1565--1576},
  publisher = {Company of Biologists},
  url = {https://journals.biologists.com/jeb/article-abstract/205/11/1565/9009},
  urldate = {2025-12-16}
}

@article{wang2000,
  title = {Vortex Shedding and Frequency Selection in Flapping Flight},
  author = {Wang, Z. Jane},
  date = {2000-05},
  year = 2000,
  journal = {Journal of Fluid Mechanics},
  volume = {410},
  pages = {323--341},
  issn = {1469-7645, 0022-1120},
  doi = {10.1017/S0022112099008071},
  url = {https://www.cambridge.org/core/journals/journal-of-fluid-mechanics/article/vortex-shedding-and-frequency-selection-in-flapping-flight/1C2EC600C1DF0CBBFBD3928CAFECAC1E},
  urldate = {2025-12-16},
  langid = {english}
}

@article{lapoMethodUnsupervisedLearning2025,
  title = {A Method for Unsupervised Learning of Coherent Spatiotemporal Patterns in Multiscale Data},
  author = {Lapo, Karl and Ichinaga, Sara M. and Kutz, J. Nathan},
  year = 2025,
  month = feb,
  journal = {Proceedings of the National Academy of Sciences},
  volume = {122},
  number = {7},
  pages = {e2415786122},
  publisher = {Proceedings of the National Academy of Sciences},
  doi = {10.1073/pnas.2415786122},
}

@article{lapoPhasorNotationDynamic2025,
Author = {Karl Lapo and Samuele Mosso and J. Nathan Kutz},
Title = {Phasor notation of Dynamic Mode Decomposition},
Year = {2025},
Eprint = {2509.03183},
journal = {arXiv},
doi = {10.48550/arXiv.2509.03183},
note = { 	
\href{https://doi.org/10.48550/arXiv.2509.03183}{https://doi.org/10.48550/arXiv.2509.03183}}
}

\end{document}